\documentclass[12pt]{iopart} 
\usepackage{iopams}

\expandafter\let\csname equation*\endcsname\relax

\expandafter\let\csname endequation*\endcsname\relax

\usepackage{amsmath}

\usepackage{amsfonts} 
\usepackage{graphicx} 

\usepackage{xcolor}

\def\half{{\textstyle \frac{1}{2}}}
\def\quarter{{\textstyle \frac{1}{4}}}

\def\d{\textrm{d} }

\DeclareMathOperator{\sech}{sech}

\begin{document}

\title[]{Geometry for evolving topographies of light-responsive plastic sheets}

\author{Mark Warner}

\address{Cavendish Laboratory, University of Cambridge, 19  JJ Thomson Avenue, Cambridge CB3 0HE, UK}
\ead{mw141@cam.ac.uk}
\vspace{10pt}
\begin{indented}
\item[]March 2019
\end{indented}

\begin{abstract}
Recently, topography change by illumination of pre-stretched, flat sheets covered in ink of optical density varying in-plane has been demonstrated by Mailen\textit{ et al}, Smart Materials and Structures, 2019. They reduce an analysis of the problem to one of metric change in the sheets in the thin limit, that is, to a question of geometry. We present the explicit form of the contraction field needed to produce the bowls these authors were interested in, using a method that can also yield the contraction field for more general desired, circularly-symmetric topography development. We give as examples the fields required for developing paraboloids and catenoids.

\end{abstract}

\vspace{2pc}

%
%
%

\section{Introduction: Topography change of plastic sheets}

In a recent paper \cite{Mailen_2019}, Mailen \textit{et al} demonstrate and analyse thermo-mechanical transformations of flat sheets to bowls. They offer a new way to achieve Gaussian curvature (GC), that is a non-trivial topography change. Nature achieves GC via differential growth \cite{Sharon2002leaves}.
GC can also be stimulated in solids  by non-uniform swelling \cite{klein2007shaping,Sharon2010,modes2011gaussian,kim2012designing}. Mailen \textit{et al} take films isotropically pre-stretched in plane before being coated with ink of spatially-varying optical density. On illumination there is accordingly a non-uniform, heat-driven, in-plane contraction, recovering to a variable extent depending on the ink density the original state before pre-stretch. Non-uniform contraction gives GC and thus slender shells of complex shape and possible mechanical strength.

To explore the essence of the route to a spherical cap, Mailen \textit{et al} model zero thickness films to avoid bend energy (in any case small for thin films) in order to concentrate on shape changes, thereby reducing the problem to the geometry of intrinsic (in-material) length changes. They explore these changes numerically. We point our (a) these changes are susceptible to simple geometrical analysis, revealing how topography is achieved, and (b) how this geometric framework can address what pattern of contractions is required to generate a general desired, cylindrically-symmetric topography.

Efrati \textit{et al}  \cite{Sharon2010} and Kim \textit{et al} \cite{kim2012designing} attack this problem of metric change directly and powerfully. They consider metrics on flat sheets that are induced to vary by having a radially-varying, in-plane swelling factor $\Omega(r)$. Some $\Omega(r)$ give surfaces with zero GC, $K = 0$, except at a point ($r=0$), that is cones. Constant positive GC surfaces, here parts of spheres, arise from an $\Omega(r) \equiv \lambda(r)^2 = c/(1 + (r/R)^2)^2$, with $K = 4/(cR)^2$, and where $\lambda$ is an in-plane contraction, $\lambda < 1$, here locally isotropic in-plane. This is the form required by Mailen \textit{et al}.

The connection between metric variation and topography is also discussed for anisotropically responsive materials with non-uniform patterning \cite{modes2010disclinations,modes2011gaussian,aharoni2014geometry}.
 The inverse problem has also been solved for cylindrically-symmetric shells \cite{aharoni2014geometry,Mostajeran2017}, where the pattern required for a $K \ne$ const. desired shape is determined. Below we employ a variant of these latter methods for these isotropic $\lambda(r)$ problems.

\section{Ink patterns for desired topographies}

\begin{figure}[h] 
\centering
\includegraphics[width=0.5\linewidth]{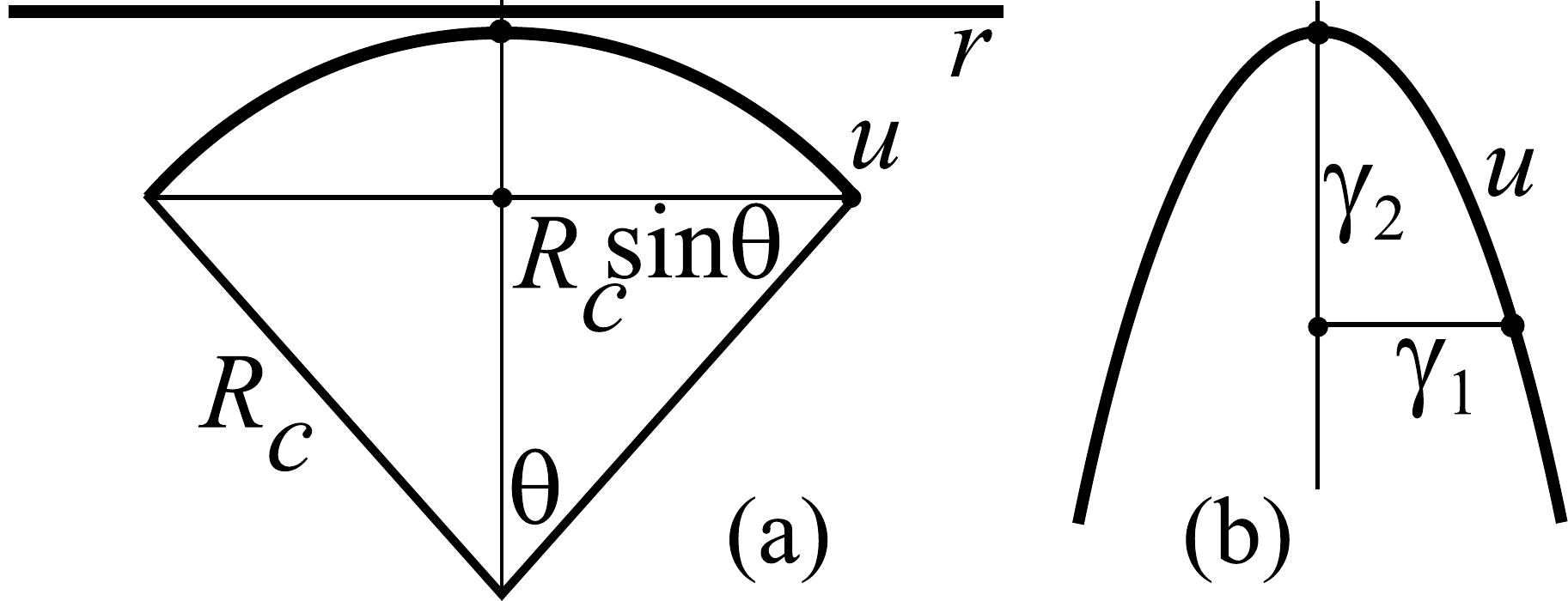} 
  \caption{ A flat disc of radius $r$ is transformed (a) into a spherical cap, seen in section, and (b) into a general shape defined by an in-space radius $\gamma_1$ and a vertical height $\gamma_2$. In each case the radius $r$ transforms into an intrinsic, in-material, radius $u$.}
  \label{fig:shells}
\end{figure}

Following the method of \cite{modes2011gaussian}, for a reference state disc suffering an isotropic contraction $\lambda(r) < 1$, dependent only on radial position $r$, one has a shrinkage of the circumference at $r$, eqn.~(\ref{eq:circ}), the contraction of a radius $r$ to an intrinsic (in-material) radius $u(r)$, eqn.~(\ref{eq:rad}), and a specification of the spherical target state by the angle $\theta$ of the position $u$, eqn.~(\ref{eq:surf}):
\begin{alignat}{4}
\lambda (r) 2 \pi r &=& 2 \pi R_c \sin\theta  &\rightarrow&  \;\; \lambda(r) r &=& R_c \sin\theta \label{eq:circ}\\
u &=& \int_0^r \d r' \lambda(r') \; &\rightarrow&  \;\;  \d u/ \d r  &=& \lambda(r) \label{eq:rad}\\
u &=& R_c \theta(u)   \;\;  &\rightarrow&  \;\; \d u/ \d \theta &=& R_c \label{eq:surf}.
\end{alignat}
where $R_c$  is the radius of curvature $K = 1/R_c^2$; see fig.~\ref{fig:shells}(a).
Equations~(\ref{eq:circ}) and (\ref{eq:rad}) involve both reference ($r$) and target ($u$) space variables, and (\ref{eq:surf}) is a specification of the target shape. The same $\lambda$ factor occurring in (\ref{eq:circ}) and (\ref{eq:rad}) shows we have the case of $\nu = -1$ of Modes \textit{et al} \cite{modes2011gaussian} where that work allows for non-isotropic shrinkage in plane.

Putting $\lambda$ of (\ref{eq:rad}) and $R_c$ of (\ref{eq:surf}) into (\ref{eq:circ}) and rearranging gives
\begin{equation}
\frac{\d \theta}{\sin\theta} = \frac{\d r}{r} \;\; \rightarrow \;\; \tan(\theta/2) = \frac{\tan(\theta_0/2)}{r_0} r \label{eq:angle}
\end{equation}
where we integrate from some inner $r_0$ and $\theta_0$ to $r$ and $\theta$.
Considering in fig.~\ref{fig:shells} very small $r_0$, we have $u_0 \simeq \lambda(0) r_0$ since $\lambda(r) \simeq \lambda(0)$, and $\theta_0 = u_0/R_c = \lambda(0) r_0 /R_c$, whence $\tan(\theta_0/2)/r_0 = \lambda(0) /2R_c$ in the limit. Then $ \tan(\theta/2) = \half  (\lambda(0)/R_c) r$. Eqn.~(\ref{eq:circ}) now leads to:
 \begin{alignat}{2}
&\lambda (r)  &=&  R_c \frac{\sin\theta}{r}  = \frac{R_c}{r} \frac{2 \tan(\theta/2)}{1 + \tan^2(\theta/2)} \nonumber\\
&\lambda (r)  &=& \frac{\lambda(0)}{1 + \left(\frac{\lambda(0)}{2}\right)^2 \left(r/R_c\right)^2} \label{eq:bowl}
\end{alignat}
in agreement with the $\lambda = \Omega^{1/2}$ result of Kim\textit{ et al} \cite{kim2012designing}, fixing $c$, $K$ and $R$ terms (the latter not to be confused with $R_c$). It is shown schematically in fig.~\ref{fig:shells}(a), and quantitatively below,  how the intrinsic radius $u$ becomes smaller than the original $r$.

\section{Surface contractions required for more general surfaces of revolution}\label{sect:general}
Fig.~\ref{fig:shells}(b) shows a more general surface $\gamma_2(\gamma_1)$ or perhaps parametrically in terms of the intrinsic radius $\gamma_1(u), \gamma_2(u)$, along with a condition for the differentiability of the surface (see eqn.~(\ref{eq:differentiability})  below). We look at the inverse problem (of which the sphere above is a trivial example): For a desired target shape, what $\lambda(r)$ is required?
Eqn.~(\ref{eq:circ}) now needs instead to connect the contracted original perimeter instead to a perimeter $2\pi \gamma_1$:
\begin{equation}
 \lambda(r) r =\gamma_1  \label{eq:circ-general}.
\end{equation}
The specification of the target surface, formerly eqn.~(\ref{eq:surf}), will now depend on the choice of target:
\begin{equation}
\gamma_2 = \half a \gamma_1^2 \quad\quad  u=\frac{1}{a}\sinh(a \gamma_1) \equiv f(\gamma_1)  \label{eq:surf-general}.
\end{equation}
Eqn.~(\ref{eq:surf-general}) specifies paraboloids and catenoids respectively for illustrations of general methods to find the corresponding $\lambda(r)$ field. An (inverse) length scale is set by $a$.
\begin{figure*}[!t] 
\centering
\includegraphics[width=0.85\linewidth]{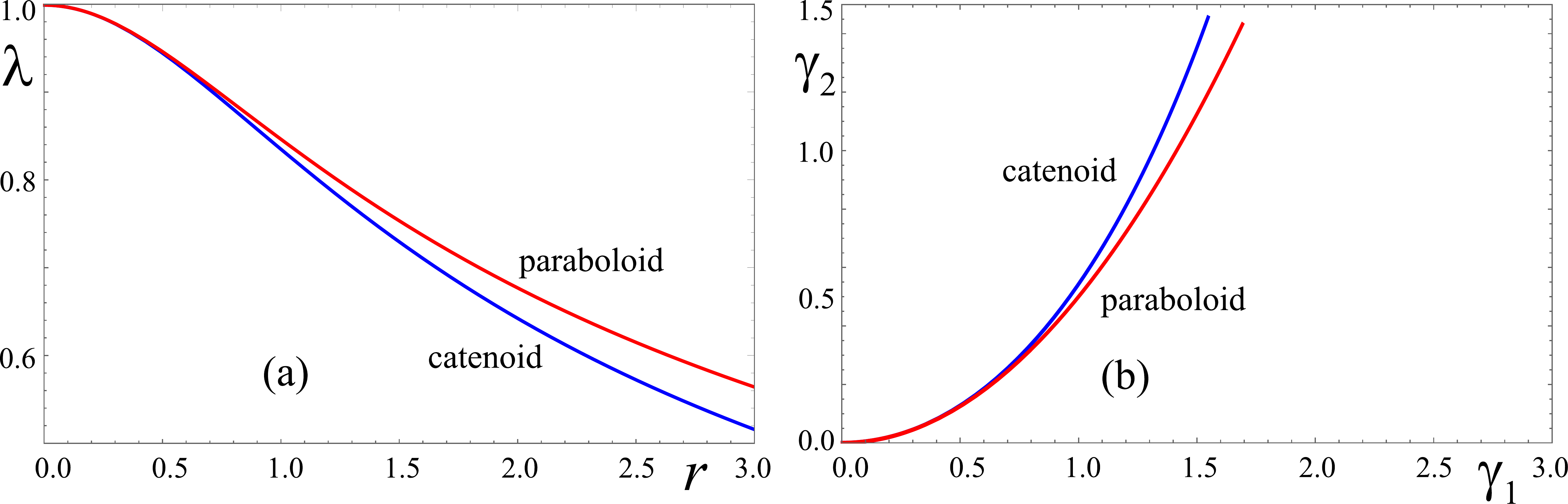}
  \caption{ A flat disc of radius $r=3$ is transformed into either a catenoid or paraboloid according to the non-uniform isotropic contractions $\lambda(r)$ given in (a). The resultant surface of revolution defined by an in-space radius $\gamma_1$ and a vertical height $\gamma_2$ is shown in each case in (b). The  intrinsic, in-material, radii $u(3)$, transformed from the initial outer value of $r=3$, are respectively 2.31 for the  paraboloid and 2.24 for the catenoid. The shrinkages reflect the accumulated contractions, eqn.~(\ref{eq:rad}).}
  \label{fig:paraboloid_catenoid}
\end{figure*}
\subsection{Paraboloids of revolution}\label{subsect:paraboloids}
The shape specifying equations, (\ref{eq:surf-general}), can be related to $r$-space: Differentiability of the surface requires
\begin{equation}
(\d u)^2 = (\d \gamma_1)^2 + (\d \gamma_2)^2 \rightarrow \d u = \d \gamma_1 (1 + a^2\gamma_1^2)^{1/2} \label{eq:differentiability}
\end{equation}
whereupon from (\ref{eq:rad}):
\begin{equation}
\lambda(r)  = \d u /\d r =\frac{ \d \gamma_1}{\d r} (1 + a^2\gamma_1^2)^{1/2} \label{eq:lambda-parabola}.
\end{equation}
Using eqn~(\ref{eq:circ-general}), that is $\gamma_1 =  r \lambda(r)$, one obtains
\begin{equation}
\d \lambda(r) /\d r = - \frac{\lambda}{r} \left(1 - 1/(1 + a^2r^2\lambda^2)^{1/2}\right) \label{eq:parabola-ODE},
\end{equation}
a non-linear, first order ODE. The quadratic coefficient $a$ is an inverse length and so it is possible to make all lengths dimensionless by absorbing $a$, for instance  $\gamma_1 \rightarrow a \gamma_1$. One must reduce not only lengths $\gamma_i$ and $u$ in the target space, but also $r$ in the reference state, otherwise the contraction factors $\lambda$ are not given properly. Without loss of generality we therefore set $a=1$ and take all lengths to be dimensionless. The same argument holds of course for catenoids too.

Eqn.~(\ref{eq:parabola-ODE})  can be solved numerically with a boundary condition $\lambda(0) = \lambda_0$, say. It is harmless about $r=0$ since an $r^2$ dependence in the numerator meets the $r$ in the denominator:
 \begin{equation*}
\d\lambda/\d r = - \frac{\lambda_0^3}{2} r + \dots \;\; \rightarrow \lambda(r \sim 0) \simeq \lambda_0 (1 - \quarter \lambda_0^2 r^2 + \dots).
\end{equation*}
For small $r$, the contraction is $\lambda(r) \simeq \lambda_0(1-\quarter \lambda_0^2 r^2 + \dots)$.  See fig.~\ref{fig:paraboloid_catenoid}(a) for the $\lambda(r)$ required for a paraboloid, while the shape (b) shows a much reduced intrinsic radius. Armed with $\lambda(r)$, shape plots are parametric in $r$, recognising $\gamma_1 = r\lambda(r)$ and $\gamma_2 = \half \gamma_1^2$.

\subsection{Catenoids of revolution}\label{subsect:catenoids}
Taking (\ref{eq:rad}) for $\lambda$ ($=\d u/\d r$), using (\ref{eq:surf-general}) for $u$, and (\ref{eq:circ-general}) for $\gamma_1$ in $\d f/\d \gamma_1$, along with the chain rule, yields:
\begin{eqnarray}
\lambda(r) &=& \cosh(r\lambda(r)) \d  (r \lambda)/\d r \nonumber\\
\d\lambda/\d r &=& - \frac{\lambda}{r} (1- \sech(r\lambda)) \label{eq:ODE-catenoid}
\end{eqnarray}
with a boundary condition $\lambda(r=0) = \lambda_0$. This ODE  is also harmless about $r=0$ since the $\sech$ in (\ref{eq:ODE-catenoid}) also vanishes when $r \rightarrow 0$: By the same argument as for paraboloids, one again obtains $\lambda(r \sim 0) \simeq \lambda_0 (1 - \quarter \lambda_0^2 r^2 + \dots)$.
Figs.~\ref{fig:paraboloid_catenoid}(a) and (b) show the resulting required $\lambda(r)$ and the catenoid, clearly with an intrinsic radius much reduced from the reference state value of $r=3$.

\section{Conclusions}\label{sect:conclusions}
The novel and versatile method of stimulating topography changes demonstrated by Mailen \textit{et al} rests ultimately on induced metric changes via a $\lambda(r)$ field of contractions. It is a process essentially determined by geometry. We have pointed out clear paths from the literature to determining the $\lambda(r)$ required to obtain the bowl shapes of interest to Mailen \textit{et al}. Further, we used an explicit and simple geometric construction to obtain the contraction field. This method is easily generalisable to determining the contraction field for the inverse problem of finding $\lambda(r)$ for a given desired topography. Two examples have been explicitly solved, paraboloids and catenoids, and these methods give paths for other embeddable shapes. The former shape, when silvered, would become an optical element on irradiative topography change. The latter is the optimal, mechanically self-supporting structure when inverted.

\section*{References}

\providecommand{\noopsort}[1]{}\providecommand{\singleletter}[1]{#1}%
\providecommand{\newblock}{}

\end{document}